\documentclass[twocolumn,trackchanges]{aastex62}
\usepackage{amsmath}
\hypersetup{
	colorlinks	= true,
	linkcolor	= red,
	urlcolor	= cyan,
	citecolor	= blue
}

\usepackage{amssymb}

\usepackage{lipsum}
\usepackage{multirow}

\graphicspath{{./images/}}

\newcommand{\exorelr}{\mbox{\textsc{ExoReL$^\Re$}}}

\turnoffediting

\accepted{August 27, 2021}

\submitjournal{AJ}

\shortauthors{Damiano \& Hu}
\usepackage{fancyhdr}
\begin{document}
	
	\title{Reflected spectroscopy of small exoplanets I: determining the atmospheric composition of sub-Neptunes planets}
	
	\correspondingauthor{Mario Damiano}
	\email{mario.damiano@jpl.nasa.gov}
	
	\author[0000-0002-1830-8260]{Mario Damiano}
	\affiliation{Jet Propulsion Laboratory, California Institute of Technology, Pasadena, CA 91109, USA}
	
	\author[0000-0003-2215-8485]{Renyu Hu}
	\affiliation{Jet Propulsion Laboratory, California Institute of Technology, Pasadena, CA 91109, USA}
	\affiliation{Division of Geological and Planetary Sciences, California Institute of Technology, Pasadena, CA 91125, USA}
	
	\begin{abstract}
		Direct imaging of widely separated exoplanets from space will obtain their reflected light spectra and measure their atmospheric properties\edit1{, and s}mall and temperate planets will be the focus of the next generation telescopes. In this work, we used our Bayesian retrieval algorithm \exorelr\ to determine the constraints on the atmospheric properties of sub-Neptune planets from observations taken with a HabEx-like telescope. Small and temperate planets may have a non-H$_2$-dominated atmosphere, and therefore, we introduced the compositional analysis technique in our framework to explore \edit1{the}\ bulk atmospheric chemistry composition without any prior knowledge about it. We have developed a novel set of prior functions for the compositional analysis free parameters. We compared the performances of the framework with the flat prior and the novel prior and we reported a better performance when using the novel priors set. 
		We found that the retrieval algorithm \edit1{can}\ not only identify the dominant gas of the atmosphere but also to constraint other less abundant gases with high statistical confidence without any prior information on the composition. The results presented here demonstrates that reflected light spectroscopy can characterize small exoplanets with diverse atmospheric composition. The Bayesian framework should be \edit1{applied}\ to design the \edit1{instrument}\ and the observation plan of exoplanet direct imaging experiments in the future.
	\end{abstract}
	
	
	\section{Introduction} \label{sec:intro}
	
	Up to date, the atmospheric exoplanet characterization of small planets, with H$_2-$dominated atmosphere, have been performed through transmission spectroscopy \edit1{(e.g., \cite{Knutson2014, Tsiaras2016B2016ApJ...820...99T, Tsiaras2019, Wakeford2017, Benneke2019a, Benneke2019b})}. However, if the atmosphere is dominated by heavier gases (e.g. H$_2$O or CO$_2$), the atmosphere is less extended and it would be more difficult to characterize it through transmission spectroscopy. \edit1{Since reflected light observations are not primarily sensitive to the atmospheric scale height, the direct imaging with reflection spectroscopy is well-suited to studying a wide range of atmospheric types including non$-$H$_2-$dominated atmosphere.} For this reason, direct imaging is gaining traction in the interest of the scientific community.
	
	A new generation of telescopes with direct-imaging capabilities are being developed to allow atmospheric characterization in the reflected light. The \textit{Nancy Grace Roman Space Telescope} (Roman, \cite{Spergel2015,Akeson2019}) will be capable of collecting starlight reflected by giant exoplanets through high-contrast imaging. \edit1{The \textit{Starshade Rendezvous Mission} (SRM, \cite{Seager2019}) could further enhance the Roman's capability to image smaller planets.} The \textit{Habitable Exoplanet Observatory} (HabEx, \cite{Gaudi2020}), a concept of a 4-m space telescope, and its main objective is to image small planets and study their atmospheres. \edit1{The \textit{Large Ultra-violet/Optical/InfraRed Surveyor} (LUVOIR, \cite{Roberge2018}) is a concept mission that would find and characterize even more Earth exoplanets similar to Earth by direct imaging.}
	
	Most of the Bayesian frameworks, used to interpret atmospheric spectra, assume that the exoplanetary atmosphere is H$_2$-dominated. The assumption simplifies the problem as only the mixing ratio of minor gases are free parameters of the model. When studying the atmospheric chemistry composition, the constraint that the sum of the mixing ratio of the gases equals to unity must be respected. The assumed knowledge of the dominant gas implicitly respect the above condition. However, small planets might not have a H$_2$-dominated atmosphere and the dominant gas is often unknown. Under these condition, compositional analysis \citep{Aitchison1982} must be introduced in the Bayesian framework. In \cite{Benneke2012}, the authors introduced compositional analysis to apply Bayesian statistics to transmission spectra of small planets. They used the centered-log-ratio (CLR) of the gases as free parameters opposed to the log-volume-mixing-ratio (Log(VMR)) as commonly used in other Bayesian framework. The authors also chose to adopt \textit{ignorance prior} (constant prior) functions for the CLR. In this work, we also adopt the CLR of gases as free parameters; however, we introduced a new set of priors. The volume mixing ratio is a direct measurement of the concentration of the gas in the atmosphere, and in our framework, we would like to explore different values of volume mixing ratio and assigning them the same probability. Since the relation between the CLR and the VMR is non-linear, re-parameterization of a constant probability distribution is not another constant distribution. We noted that with the new set of priors, the Bayesian framework does not overestimate or underestimate the concentration of gases, which occurred when we used a constant prior function for CLR.
	
	Here we use a robust Bayesian inverse retrieval method \citep[\exorelr,][]{Damiano2020a, Damiano2020b} to interpret reflected light spectra of small and temperate sub-Neptune planets. This is, to our knowledge, the first time a retrieval method is applied to reflected light spectroscopy of small exoplanets beyond Earth-like planets \citep[which was discussed in ][]{Feng2018}. Our model has the following specific advancements. (1) we can use retrieval to determine the main component of the atmosphere, rather than assuming the atmosphere to be, for example, N$_2$-dominated or H$_2$-dominated. (2) We can explore atmosphere in which multiple gases have equal or similarly dominant abundances. (3) we can retrieve the mixing ratio of H$_2$O below a water cloud, leading to potential inference of an ocean.
	
	The paper is organized as follows: Sec.~\ref{sec:model} describes the setup of the retrieval,  Sec.~\ref{sec:scenarios} discusses the scientific scenarios we explored in this work, Sec.~\ref{sec:result} shows the results of the inverse retrieval performed on the scientific scenarios, and Sec.~\ref{sec:discussion} explores the implications of the results obtained and the retrieval setup used.
	
	\section{Methods} \label{sec:model}
	
	\subsection{Retrieval Setup} \label{sec:retrieval}
	
	We update the atmospheric scenario and radiative transfer model in \cite{Hu2019B2019ApJ...887..166H,Damiano2020a,Damiano2020b} to enable simulations of super-Earths' reflected light spectra. Compared with giant planet models, a key difference for super-Earths' atmospheres is that they can have \edit1{a wider range of bulk atmospheric compositions} \cite[e.g.,][]{Hu2014B2014ApJ...784...63H}. In order to consistently calculate the volume mixing ratio (VMR) of the constituents of an exoplanetary atmosphere, models must respect the following relations:
	
	\begin{equation}
	    0 < VMR_i < 1
	    \label{eq:vmr1}
	\end{equation}
	
	and 

    \begin{equation}
        \sum_{i=1}^{n} VMR_i = 1
        \label{eq:vmr2}
    \end{equation}
	
	\noindent where $n$ represents the total number of gases considered in the model. 
	The equations above underline that the values of the gases are not independent as one value can be calculated by knowing the other $n-1$ values. Therefore, in the gaseous giant planets scenario, where H$_2$ is always the dominant gas, the Bayesian statistical process uses the \textit{Log(VMR)} of the other gases as free parameters. This scheme always ensure that Eq. \ref{eq:vmr1} and \ref{eq:vmr2} are respected. In the case of smaller planets, it is unknown which gas is the dominant one a priori. For this reason, it is challenging using the \textit{Log(VMR)} as free parameter for each of the considered gases in the Bayesian process while respecting Eq. \ref{eq:vmr1} and \ref{eq:vmr2}. 
	
	To overcome this difficulty, we implemented the centered-log-ratio (CLR) \citep{Aitchison1982} of the mixing ratios of H$_2$O, CO$_2$, CH$_4$, N$_2$, O$_2$, and O$_3$ as free parameters. The CLR is defined for a composition \textit{\textbf{X}} of $n$ elements as follows:
	
	\begin{equation}
	    clr(\textbf{X}) = ln\left[ \frac{X_i}{g_m(\textbf{X})}, ..., \frac{X_n}{g_m(\textbf{X})} \right]
	    \label{eq:clr2}
	\end{equation}
	
	\noindent where
	
	\begin{equation}
	    g_m(\textbf{X}) = \left( \prod_{i=1}^{n} X_i \right)^{1/n}
	    \label{eq:clr3}
	\end{equation}
	
	\noindent is the geometric mean of the composition \textit{\textbf{X}}.
	
	\edit1{Let us define the \textbf{VMR} as a composition of $n$ gases. By using the CLR, the restrictions the model must respect can be simplified.} Indeed, Eq. \ref{eq:vmr1} and \ref{eq:vmr2}, in the CLR space, simply translate into:
	
	\begin{equation}
        \sum_{i=1}^{n} clr(\textbf{VMR})_i = 0
        \label{eq:clr1}
    \end{equation}
    
    \noindent By definition, the CLR can assume positive and negative values and in particular the CLR could assume any values inside the range (-$\infty$, +$\infty$). Lower values of CLR correspond to low value of VMR and vice versa. One can thus specify $n-1$ CLR values and derive the last CLR value without favoring or disfavoring any species. This approach for compositional retrieval has been previously applied to transmission spectroscopy \citep{Benneke2012}.
	
	In this study, we include water clouds as the only type of condensates in the atmosphere. This is consistent with the current plans for direct imaging missions. We use the optical properties from \cite{Palmer1974} to calculate the cross-sections and single scattering albedo of water droplets. In this way, our model includes the absorption bands of water clouds at 1.44 $\mu$m and 1.93 $\mu$m, and gradually more absorption of the clouds outside the bands when the wavelength is longer than $\sim1$ $\mu$m. The description of the cloud follows \cite{Damiano2020a}, which preserves the causal relationship between the condensation of water and the formation of the cloud. In our model, the cloud is parametrized by the cloud top pressure ($P_{\rm top}$), the cloud depth (i.e., the difference in pressure between the cloud bottom and the cloud top, $D_{\rm cld}$), and the condensation ratio (i.e., the ratio between the mixing ratio of water above the cloud and that below the cloud, denoted as CR).
	Finally, instead of calculating the pressure-temperature profile, we assume the atmosphere to be isothermal and assign an estimated temperature given the irradiation level. This greatly simplifies the forward model and is valid because the reflected light spectra are not sensitive to the specifics of the temperature \citep{Feng2018,Robinson2018}. 
	
	In the vision of performing the retrieval of super-Earth reflected light spectroscopy, we introduce the surface pressure and albedo as fixed or free parameters of the model. Even though they are not leveraged by the scenarios we present here (in this work they are fixed to a specific value) (see Sec. \ref{sec:scenarios}), we will describe and explore the effect of the surface (pressure position and ground albedo) in a subsequent paper on small planets (Reflected spectroscopy of small planets II, Damiano \& Hu in prep.).
	
	\subsection{A new set of priors} \label{sec:priors}
	
	In \exorelr\  \citep{Hu2019B2019ApJ...887..166H,Damiano2020a,Damiano2020b}, we defined the $Log(VMR)$ of H$_2$O, NH$_3$, and CH$_4$ as free parameters of our Bayesian model to retrieve insights into the atmosphere of gaseous giant planets. We used uniform priors for the $Log(VMR)$ to assign the same probability to the values in the range [-12.0, 0.0). To characterize small planets, we pointed to the CLR of the gases to be the free parameters of the Bayesian process. However, the CLR is a non-linear transformation of the VMR space. For this reason, we can not use a uniform set of priors for the CLR, otherwise we would assign different probabilities to different values of VMR, favoring some VMR values rather than equally sampling the whole range. This is particularly true for values of VMR close to unity, this leads the gas with the strongest absorption features to be misinterpreted as the dominant gas in the atmosphere.
	
	For this reason, in this work, we introduce a new set of priors for the CLR. First of all, we underline that the number of free parameters related to gases is equal to $n-1$ with $n$ being the total number of gases. This is the result of Eq. \ref{eq:clr1}. Therefore, we define a `filler' gas, and in our model H$_2$ is always considered as the `filler' given its nature of inactive gas (i.e. it does not show spectral features in the moderate resolution and S/N relevant to exoplanet observations). The choice of the filler gas does not favor or disfavor that gas by itself.
	
\begin{figure*}
	    \centering
	    \includegraphics[width=\textwidth]{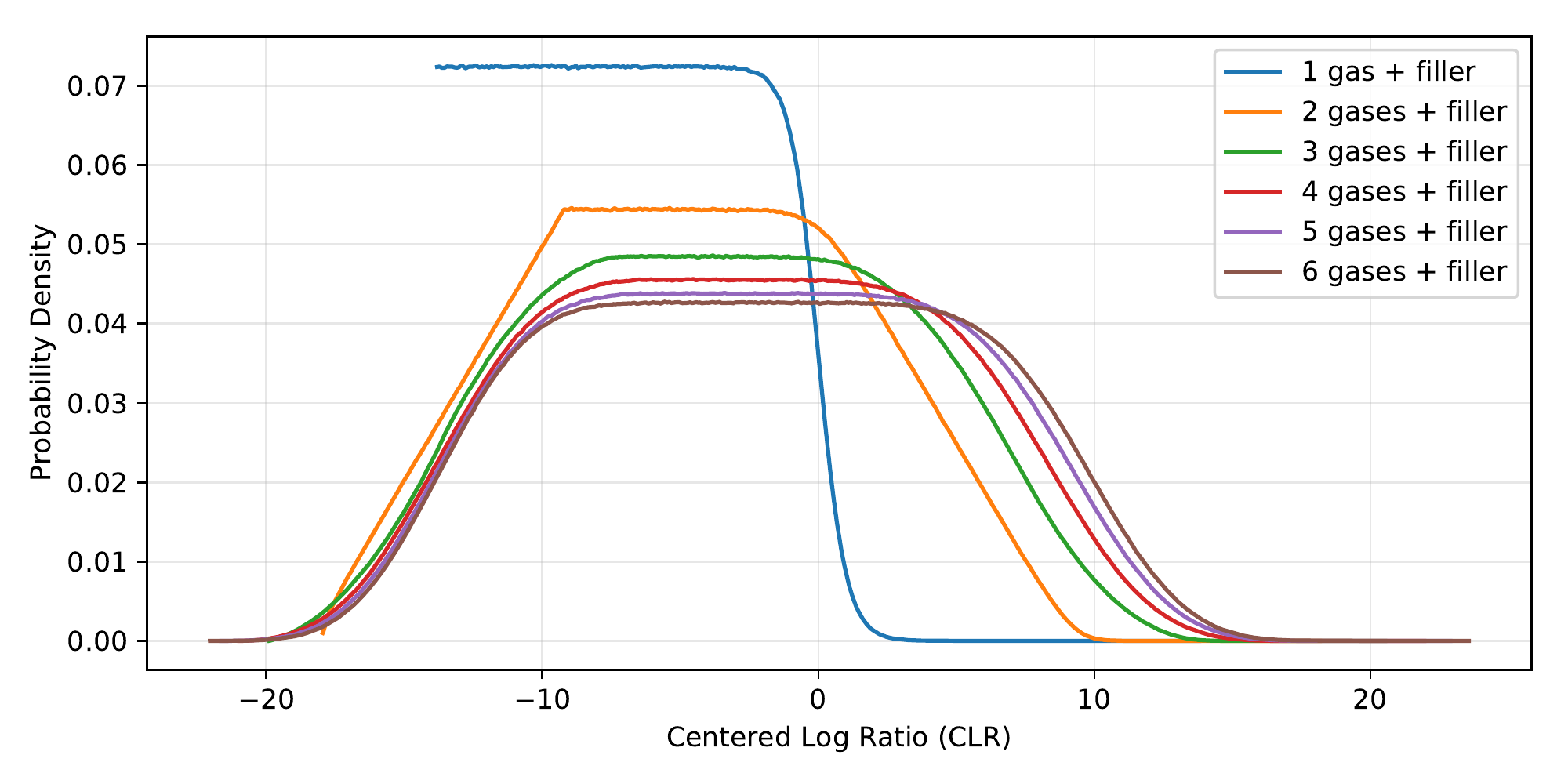}
	    \caption{Prior functions in the centered log ratio space for different number of gases considered in the retrieval. The functions are normalized by their integral.}
	    \label{fig:priors2}
	\end{figure*}

	Let us consider a VMR range defined within [10$^{-12}$, 1). For each $n-1$ VMR$_i$, we randomly draw a value from the range according to an \textit{uniform distribution}. We keep only those sets of VMR where the sum is less than 1, therefore we can include the dependent VMR of the filler gas to sum to unity. We use Eq. \ref{eq:clr2} to transform the sets of VMR into sets of CLR. By doing so we transform an \textit{uniform distribution} defined in the VMR space into a distribution of the CLR space. Since the CLR is not a linear transformation, the distribution defined in the CLR space is not uniform. Moreover, we observed that the distribution in the CLR space depends on the value of $n$, meaning that the shape of the CLR prior is different for different number of gases to be fit in the Bayesian process. In Fig. \ref{fig:priors2}, we reported the resulting probability density function for the CLR for different number of gases considered in the Bayesian process to be fit. Note that once the number $n$ of gases has been chosen, the same prior function is used for all the $n$ gases.
	
	Finally, the prior functions associated to all the other free parameters, i.e. surface pressure P$_0$, cloud top pressure P$_{top}$, the cloud depth $D_{cld}$, condensation ratio CR, the surface albedo A$_g$, and the Log of the gravity Log(g), have uniform distributions as introduced in \citep{Damiano2020a}.
	
	\section{Planetary Scenarios} \label{sec:scenarios}
	
	\subsection{Idealized scenarios}
	
\begin{figure}[!h]
	    \centering
	    \includegraphics[width=\columnwidth]{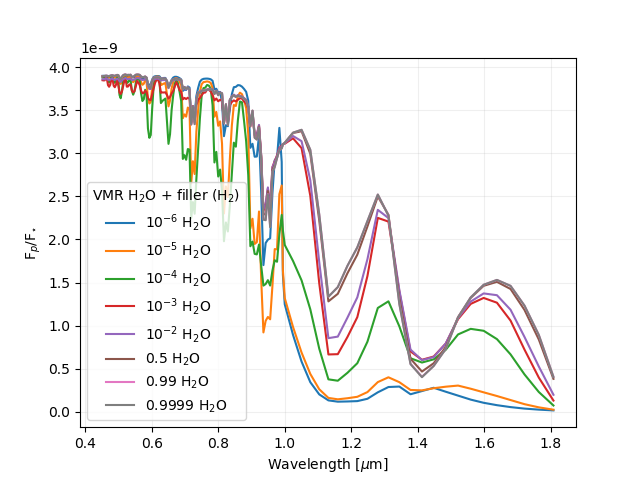}
	    \caption{Reflected light spectra of a hypothetical H$_2$-H$_2$O atmosphere with different amount of water. When water is below 10$^{-2}$ in volume mixing ratio the spectra are distinguishable, while above 10$^{-2}$ the spectra are degenerate and not sensitive anymore to the volume mixing ratio of the two gases.}
	    \label{fig:wh}
	\end{figure}

	To test the performance of the new \exorelr's set up, we include here the analysis of the simplest scenario possible, i.e. an atmosphere made solely of water and hydrogen. We simulated 8 spectra to cover the VMR of water and hydrogen between 10$^{-6}$ and 0.999999. Fig. \ref{fig:wh} shows the simulated flux ratio of a 10 M$_\oplus$ and 2.4 R$_\oplus$ planet with a gravitational acceleration of $g=17$ m/s$^2$ around a Sun-like star at 1 AU with different atmospheric compositions. It is worth noting before running the statistical retrieval process that when the amount of water is significant ($\ge$10$^{-2}$) the spectra are degenerate and can not be discerned from each other. When the water amount is low, the water cloud layer is thin and the starlight passes through. On the contrary, when the water concentration is higher, the clouds become optically thick and the modulation observed in the flux ratio is mostly due to the water cloud single scattering albedo. Note that in all the simulations the cloud position has been kept the same around 10$^5$ Pa. 
	
	Although the spectra are degenerate when the water mixing ratio is $>10^{-2}$, the mean molecular mass and therefore the scale height of the atmosphere is different between, for example, 1\% water and 50\% water. This difference can be useful to break the degeneracy when other absorbing gases are included (e.g. CO$_2$ and CH$_4$), because the atmospheric scale height will affect the depth of the absorption features of the other absorbing gases (as shown in the realistic scenarios).
	
    \subsection{Realistic scenarios}
	
	For more complex scenarios, we simulate the same type of planet as in the idealized scenario, i.e.
	10 M$_\oplus$, 2.4 R$_\oplus$, g=17.0 m s$^{-2}$, around a Sun-like star at 1 AU, but with different atmospheric compositions. This type of planet is close to the center of the high-radius population of the 1.7$-$3.5 R$_\oplus$ planet population found by the Kepler survey \citep{Fulton2018}. Such a planet can be a rocky core with an extended H$_2$-dominated atmosphere (i.e., a gas dwarf), or a planet with 50\% of mass being water (i.e., a water world) \citep{Zeng2019}. If it is a water world, the extended H$_2$-dominated atmosphere is not required to explain the radius, and it can have a thin H$_2$-dominated atmosphere, or an N$_2$- or CO$_2$-dominated atmosphere. This reduces to determining the mean molecular weight of the atmosphere. 
	
\begin{figure*}[!ht]
	    \centering
	    \includegraphics[width=\textwidth]{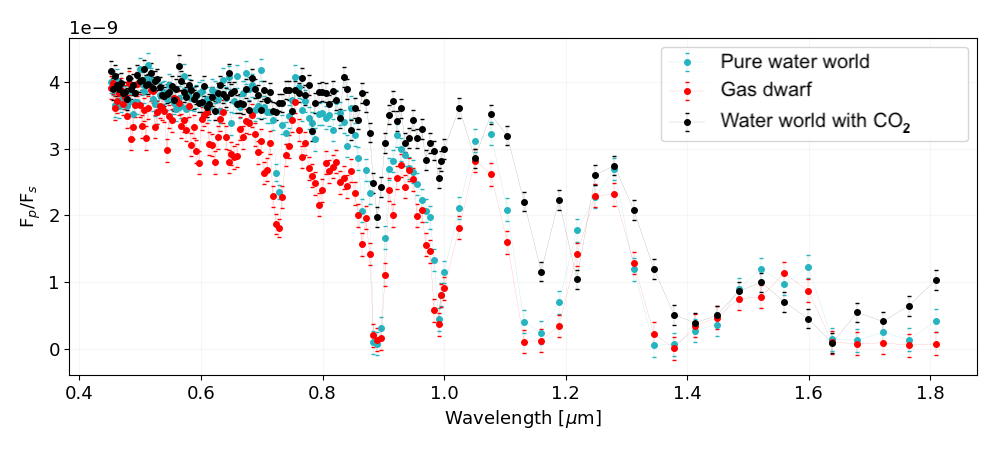}
	    \caption{Simulated reflected spectra of the three scenarios of sub-Neptune exoplanets.}
	    \label{fig:scenarios}
	\end{figure*}

	We include three different scenarios to cover these possibilities. 1) pure water world (Fig. \ref{fig:scenarios}, cyan model): in this scenario we have a high concentration of water below the water cloud layer. On top of the cloud, instead, hydrogen is the dominant gas as the water condensed in the cloud form. We also included a small concentration of CO$_2$, CH$_4$ and N$_2$. 2) gas dwarf (Fig. \ref{fig:scenarios}, red model): this scenario represent the sub-Neptune class of planets with hydrogen dominated atmosphere, a water cloud layer and absorbing gas (e.g., CO$_2$, CH$_4$ and N$_2$). 3) water world with CO$_2$ (Fig. \ref{fig:scenarios}, black model): in this scenario we will test the Bayesian retrieval in an environment in which there are two dominant gases (i.e., H$_2$O and CO$_2$) of similar abundance, a water cloud layer and absorbing gas (i.e., CH$_4$) and spectrally inactive gases (i.e., N$_2$ and H$_2$).
	
	As we can note from Fig. \ref{fig:scenarios}, the three models have the same baseline in terms of flux ratio because the physical parameters of the planet and the simulated system have not been changed. The differences in the spectra are only due to different chemical composition of the simulated atmosphere. The three spectra show some similarities but the main constituent is different in each of them. The atmospheric values used to simulate the three spectra are reported in Tab. \ref{tab:sims}.
	
	For all the scenarios presented here, we considered a signal to noise ratio (S/N) of 20 at 0.7 $\mu m$ at a spectral resolution of 140 in the optical ($\lambda$ $\leq$ 1.0 $\mu m$) and 40 in the infrared ($\lambda$ $\geq$ 1.0 $\mu m$) without making any specific assumption on the integration time which is not the focus of this work. This is consistent with the science plan of HabEx \citep{Gaudi2020}.
	
	\begin{deluxetable*}{cccc}
		\tablecaption{Atmospheric parameters used to simulate the realistic scenarios. \label{tab:sims}}
		\tablehead{
			\colhead{Parameter} & \colhead{Pure water world} & \colhead{Gas dwarf} & \colhead{Water world with CO$_2$}}
		\startdata
		$Log(P_{top, H_2O})$ [Pa] & $4.85$ & $4.85$ & $4.85$ \\
		$Log(D_{cld, H_2O})$ [Pa] & $4.30$ & $4.30$ & $4.30$ \\
		$Log(CR_{H_2O})$ & $-5.00$ & $-5.00$ & $-5.00$ \\
		$Log(VMR_{H_2O})$ & $-0.30$ & $-3.00$ & $-0.35$ \\
		$Log(VMR_{CO_2})$ & $-3.40$ & $-3.40$ & $-0.35$ \\
		$Log(VMR_{CH_4})$ & $-3.40$ & $-3.40$ & $-3.40$ \\
		$Log(VMR_{N_2})$ & $-3.40$ & $-3.40$ & $-3.40$ \\
		$g\ [m/s^2]$ & $17.0$ & $17.0$ & $17.0$ \\
		$\mu$ & $10.05$ & $2.06$ & $28.13$ \\
		\enddata
	\end{deluxetable*}
	
	\section{Results} \label{sec:result}
	\subsection{Idealized scenario} \label{sec:idialized}
	
	As we described in Sec. \ref{sec:scenarios}, we initially tested our updated model with the simplest idealized case: an atmosphere with solely water and the filling gas, i.e., H$_2$. We run 8 retrievals as the simulated cases in which the amount of water and hydrogen in the atmosphere was changed. When the mixing ratio of water (below the cloud) is in the range of $10^{-5}\sim10^{-3}$, the cloud position and the water abundance below the cloud can be retrieved (Fig.~\ref{fig:S_ideal}). As shown in the spectra (Fig.~\ref{fig:wh}), when the volume mixing ratio of water is larger than 10$^{-2}$ the spectra are degenerate and the model is not able to correctly identify the truths (i.e., producing a broad posterior distribution of water mixing ratio between $10^{-2}$ and 1, Fig.~\ref{fig:S_ideal}). The retrieval model always identifies H$_2$ as the dominant gas even though the role of the two gases swaps in the 8 simulations (Fig.~\ref{fig:S_ideal}). The major driver of these degeneracies is the water cloud in the atmosphere. When the water amount is is low, the clouds are not completely opaque in all layers, and therefore some of the light can reach deep into the cloud layers of the atmosphere. On the contrary, when the VMR of water is higher than 10$^{-2}$ the cloud becomes optically thick and behaves more like a reflecting deck, causing the spectrum to be dominated solely by the cloud single-scattering albedo.
	
	Using the idealized scenarios we have also tested the performance of the new prior functions developed in Section~\ref{sec:priors} and the flat prior in the CLR space. Fig.~\ref{fig:S_flat_vs_mod} shows the posterior distribution calculated by \exorelr\ when the two set of priors are used to retrieve one of the idealized cases, i.e. 10$^{-3}$ H$_2$O scenario. In the case of flat prior, the framework found two solutions. One of the results is consistent with the true value, the other result shows different atmospheric characteristics. In particular, the retrieved amount of water is close to unity, and the cloud layers extend deeper in the atmosphere.
	If the modified priors are used (blue model in Fig. \ref{fig:S_flat_vs_mod}), only one solution is found and it is the one consistent with the truths.
	
	\subsection{Water world}
	
	\begin{deluxetable}{cccc}[!h]
		\tablecaption{Atmospheric retrieved parameters for the Water World scenario. The parameters that are usefully constrained are checked.  \label{tab:sc1_res}}
		\tablehead{
			\colhead{Parameter} & \colhead{3$\sigma$} & \colhead{2$\sigma$} & \colhead{1$\sigma$}}
		\startdata
		$Log(P_{top, H_2O})$ & $\checkmark$ & $\checkmark$ &  \\
		$Log(D_{cld, H_2O})$ & $\checkmark$ & $\checkmark$ &  \\
		$Log(CR_{H_2O})$ & $\checkmark$ &  &  \\
		$Log(VMR_{H_2O})$ &  &  &  \\
		$Log(VMR_{CO_2})$ & $\checkmark$ & $\checkmark$ & $\checkmark$ \\
		$Log(VMR_{CH_4})$ & $\checkmark$ & $\checkmark$ & $\checkmark$ \\
		$Log(VMR_{N_2})$ & $\checkmark$ & $\checkmark$ & $\checkmark$ \\
		$Log(VMR_{H_2})^{\star}$ &  &  &  \\
		$g\ [m/s^2]$ &  &  &  \\
		$\mu^{\star}$ &  &  &  \\
		\enddata
	\end{deluxetable}
	
	The first of the three realistic scenarios we considered in this work is a water world. In this scenario, the amount of water below the clouds is considered to be high, then it drops with altitude as it condenses into cloud form.
	
	\begin{figure}[!ht]
	    \centering
	    \includegraphics[width=\columnwidth]{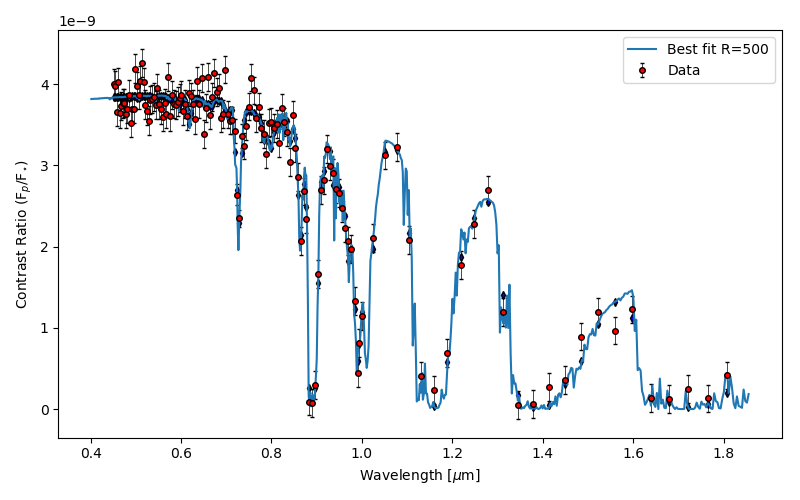}
	    \caption{Simulated data and retrieved model in the water world scenario}
	    \label{fig:scen_s1}
	\end{figure}
	
	The simulated data and the best fit calculated by \exorelr\ are shown in Fig. \ref{fig:scen_s1}. Our model was able to fit the data, and in particular it was able to reconstruct all the spectroscopic features due to the absorption of CO$_2$ and CH$_4$. In the Appendix \ref{sec:A1} we reported the 1D probability distribution functions of the free and derived parameters along the 2D correlation maps of the parameters. In the posterior distribution (Fig. \ref{fig:s1}) the 2$\sigma$ confidence level is reported. In Tab. \ref{tab:sc1_res}, we indicated whether or not the true value of the free parameters falls inside 1, 2, or 3$\sigma$ of the posterior distribution in Fig. \ref{fig:scen_s1}. The result calculated by \exorelr\ (see Fig. \ref{fig:s1}) gives a clear overview of the performances of the model in handling cases with completely unknown atmospheric composition. First of all, the model was able to correctly retrieve the position of the water cloud in the atmosphere (P$_{top,H_2O}$ and D$_{H_2O}$) within 2$\sigma$. Secondly, the overall composition of the atmosphere has been retrieved with CH$_4$, CO$_2$, and N$_2$ retrieved within 1$\sigma$. The amount of water, and therefore the concentration of hydrogen (derived), was not retrieved within 3$\sigma$. However, the model was able to identify water as the dominant gas below the cloud layer, and hydrogen dominated above it. This can be easily visualized on the retrieved chemistry profile of the atmosphere reported in Appendix \ref{sec:A1}. Because of the over-estimation of the water content in the atmosphere, the fitted gravity of the planet and the mean molecular mass ($\mu$, derived) are also slightly off.
    
    Finally, it is worth noting that the model was able to identify the atmosphere to be a water-dominated one. In comparison, the model applied to the same atmosphere without CH$_4$ or CO$_2$ (i.e., the 50\% water case in the idealized scenarios, Sec.~\ref{sec:idialized}) would return a water mixing ratio between $10^{-2}$ and unity without preference, and thus cannot identify the atmosphere being hydrogen or water-dominated. Including other absorbing gases breaks the degeneracy. The absorption features of CH$_4$ and CO$_2$ vary in depth accordingly to the thickness of the atmosphere. If the atmosphere was lighter, i.e., an H$_2-$dominated atmosphere, the absorption features would have been deeper.
	
	\subsection{Gas dwarf}
	
	\begin{deluxetable}{cccc}[!h]
		\tablecaption{Atmospheric retrieved parameters for the Sub-Neptune scenario. The parameters that are usefully constrained are checked. \label{tab:sc2_res}}
		\tablehead{
			\colhead{Parameter} & \colhead{3$\sigma$} & \colhead{2$\sigma$} & \colhead{1$\sigma$}}
		\startdata
		$Log(P_{top, H_2O})$ & $\checkmark$ & $\checkmark$ & $\checkmark$ \\
		$Log(D_{cld, H_2O})$ & $\checkmark$ & $\checkmark$ & $\checkmark$ \\
		$Log(CR_{H_2O})$ & $\checkmark$ & $\checkmark$ &  \\
		$Log(VMR_{H_2O})$ & $\checkmark$ & $\checkmark$ & $\checkmark$ \\
		$Log(VMR_{CO_2})$ & $\checkmark$ & $\checkmark$ & $\checkmark$ \\
		$Log(VMR_{CH_4})$ & $\checkmark$ & $\checkmark$ &  \\
		$Log(VMR_{N_2})$ & $\checkmark$ & $\checkmark$ & $\checkmark$ \\
		$Log(VMR_{H_2})^{\star}$ & $\checkmark$ & $\checkmark$ & $\checkmark$ \\
		$g\ [m/s^2]$ & $\checkmark$ & $\checkmark$ & $\checkmark$ \\
		$\mu^{\star}$ & $\checkmark$ & $\checkmark$ & $\checkmark$ \\
		\enddata
	\end{deluxetable}
	
	In this scenario, we simulated a small planet with an H$_2-$dominated atmosphere with a water cloud layer and other absorbing gases.
	
	\begin{figure}[!ht]
	    \centering
	    \includegraphics[width=\columnwidth]{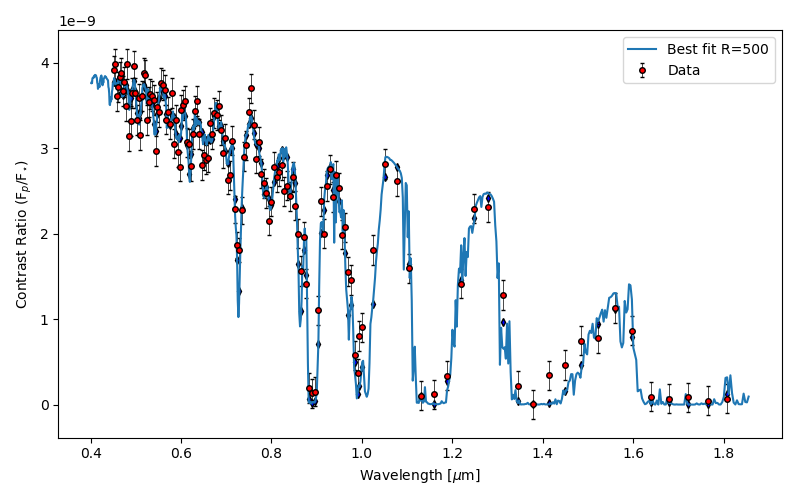}
	    \caption{Simulated data and retrieved model in the gas dwarf scenario}
	    \label{fig:scen_s2}
	\end{figure}
	
	The simulated data and the best fit calculated by \exorelr\ are shown in Fig. \ref{fig:scen_s2}. The model was able to fit the data, and to recover all the spectroscopic features due to the absorption of CO$_2$ and CH$_4$.
	In the Appendix \ref{sec:A2} we reported the full posterior distribution calculated by \exorelr\ (see Fig. \ref{fig:s2}). In Tab. \ref{tab:sc2_res}, we indicated whether or not the true value of the free parameters falls inside 1, 2, or 3$\sigma$ of the posterior distribution in Fig. \ref{fig:scen_s2}. The result (see Fig. \ref{fig:s2}) gives a clear overview of the performances of the model in handling cases in which the dominant gas is a spectrally inactive gas. Also in this case, the model was able to correctly retrieve the position of the water cloud in the atmosphere (P$_{top,H_2O}$ and D$_{H_2O}$) within 1$\sigma$. The overall composition of the atmosphere has been retrieved within 1$\sigma$ including water and hydrogen. This can be easily visualized on the retrieved chemistry profile of the atmosphere reported in Appendix \ref{sec:A2}. In this case, the gravity of the planet and the derived mean molecular mass match the simulated values within 1$\sigma$.
	
	\subsection{Water world with CO$_2$}
	
	\begin{deluxetable}{cccc}[!h]
		\tablecaption{Atmospheric retrieved parameters for the H$_2$O$-$CO$_2$ atmosphere scenario. The parameters that are usefully constrained are checked. \label{tab:sc3_res}}
		\tablehead{
			\colhead{Parameter} & \colhead{3$\sigma$} & \colhead{2$\sigma$} & \colhead{1$\sigma$}}
		\startdata
		$Log(P_{top, H_2O})$ & $\checkmark$ & $\checkmark$ &  \\
		$Log(D_{cld, H_2O})$ & $\checkmark$ & $\checkmark$ & $\checkmark$ \\
		$Log(CR_{H_2O})$ & $\checkmark$ & $\checkmark$ &  \\
		$Log(VMR_{H_2O})$ & $\checkmark$ & $\checkmark$ & $\checkmark$ \\
		$Log(VMR_{CO_2})$ & $\checkmark$ & $\checkmark$ & $\checkmark$ \\
		$Log(VMR_{CH_4})$ & $\checkmark$ & $\checkmark$ &  \\
		$Log(VMR_{N_2})$ & $\checkmark$ & $\checkmark$ & $\checkmark$ \\
		$Log(VMR_{H_2})^{\star}$ & $\checkmark$ & $\checkmark$ & $\checkmark$ \\
		$g\ [m/s^2]$ & $\checkmark$ & $\checkmark$ & $\checkmark$ \\
		$\mu^{\star}$ & $\checkmark$ & $\checkmark$ & $\checkmark$ \\
		\enddata
	\end{deluxetable}
	
	In this last case, we wanted to simulate an heavy atmosphere without having a single dominant gas. Below the cloud the concentrations of H$_2$O and CO$_2$ are $\sim$0.5 of the total atmosphere. Above the cloud, the water abundance dropped as it condensed, while the volume mixing ratio of CO$_2$ was kept constant. 
	
	\begin{figure}[!ht]
	    \centering
	    \includegraphics[width=\columnwidth]{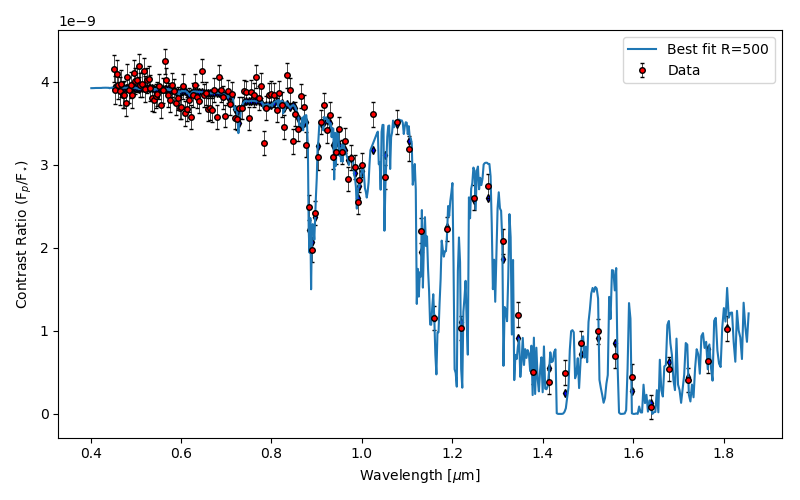}
	    \caption{Simulated data and retrieved model in the water world with CO$_2$ scenario}
	    \label{fig:scen_s3}
	\end{figure}
	
	This is the case where the compositional analysis shows its strength. The simulated data and the best fit calculated by \exorelr\ are shown in Fig. \ref{fig:scen_s3}. Also in this case, the model calculated by \exorelr\ fits the simulated data. Moreover, also the free parameters have been retrieved with high confidence. In Tab. \ref{tab:sc3_res}, we indicated whether or not the true value of the free parameters falls inside 1, 2, or 3$\sigma$ of the posterior distribution in Fig. \ref{fig:scen_s3}. We reported the full posterior distribution in the Appendix \ref{sec:A3}. The cloud has been positioned with high confidence at 10$^5$ Pa, and the gravity of the planet was correctly estimated.
	
	The important result here is that the model was able to identify both CO$_2$ and H$_2$O as dominant gases without showing degenerate solutions. The overall composition has been retrieved correctly within 1$-$2 $\sigma$. It is worth nothing that the concentration of N$_2$ has a broad probability distribution, suggesting that there might no be enough information to exactly fit it. Also, the active gases (i.e. H$_2$O, CO$_2$, and CH$_4$) show correlations with each other within a small range of values therefore not impacting the overall solution. As the other cases, the retrieved chemical profile is provided (see Appendix \ref{sec:A3}). From here, it is possible to easily visualize that \exorelr\ was able to correctly interpret the chemical structure of the atmosphere.
	
	\section{Discussion} \label{sec:discussion}
	
	In the work presented here, we updated our model \exorelr\ to interpret exoplanetary atmospheres through compositional analysis without providing any prior information about the dominant gas. This is an important step to achieve as small planets (e.g., Sub-Neptunes, Super-Earths, Earth-like planets) might not have H$_2$ as dominant constituent of their atmosphere. 
	
	\subsection{The set of priors}
	
	We introduced in our model the centered log ratio (CLR) of the gases as free parameter. The CLR is a more robust tool for compositional analysis compared to the volume mixing ratio (VMR) which is usually used for Bayesian inverse retrieval processes, because it does not favor or disfavor the ``filling gas''. The CLR and the VMR are related to each other in a non-linear relation (Eq. \ref{eq:clr2}). However, we are interested in exploring the VMR as it is used within the radiative transfer model and it is a direct way to asses the composition of the atmosphere. Therefore, we developed a novel set of prior functions for the CLR within the Bayesian statistical process. This was necessary since we wanted to explore the VMR in a range within [10$^{-12}$, 1) assigning the same probability to the values within the range. Since the relation between CLR and VMR is non-linear the resulted prior functions for the CLR are non-constant across the range (Fig. \ref{fig:priors2}).
	
	The resulted shape of the CLR prior functions prevents the abundance for a gas to be extremely high or extremely low unless the data suggests otherwise. However, this does not have an impact in the ability of the model to identify the dominant gas in the atmosphere as suggested by the studied scenarios presented in this work. In Fig. \ref{fig:S_flat_vs_mod}, we showed that in the case where the framework with the flat prior, a solution with high concentration of water was found. This solution was not found when the modified prior functions were used within the statistical calculation.
	
	\subsection{The H$_2$O - H$_2$ degeneracy}
	
	When we applied our updated model to a simple idealized case, we found degenerate solutions. In particular, when only water and hydrogen were included in the atmosphere, the model found degenerate solutions when the VMR of water was larger than 10$^{-2}$. As we explained in the Result section, since \edit1{hydrogen} does not have strong spectral features in the wavelength range probed in this work (0.4 $-$ 1.8 $\mu m$), when the amount of water is high, the water cloud that forms is dense enough to be optically thick, and the reflected spectrum is dominated by the single scattering albedo of the cloud. Otherwise, if the water content in the atmosphere is between 10$^{-5}$ $\sim$ 10$^{-3}$ the Bayesian process is able to correctly retrieve the VMR of water and hydrogen.
	
	As we have shown with the water world scenarios with and without CO$_2$, the degeneracies might be broken when spectrally active gases are included in the model. This is because the spectral signatures are affected by the scale height of the atmosphere, and changing the H$_2$O/H$_2$ ratio will affect the mean molecular mass ($\mu$) which will ultimately change the atmospheric scale height.
	
	\subsection{Atmospheric scenarios}
	
	We applied our model to different atmospheric scenarios of cold sub-Neptunes planets. The model demonstrated that it will be possible to identify the dominant gas of the atmosphere (even if it is not hydrogen) and the overall composition with high confidence without any prior knowledge. 
	In the case of the gas dwarf scenario, the framework was able to determine that the atmosphere is H$_2$-dominated, to retrieve with high confidence the composition of the atmosphere, and to place correctly the water cloud in the atmosphere.
	
	The framework also demonstrated that it would be able to interpret an atmosphere in which there are multiple spectrally active dominant gases. In the case of the water world with CO$_2$ case, the retrieved values of the VMR of the gases are within 1-2$\sigma$. The water cloud position has also been retrieved within 1$\sigma$. We also reported that the framework did not perform optimally in case of two dominant gases in which one of the two is a spectrally inactive gas, e.g. pure water world scenario. In that case even though the two dominant gases have been identified, their concentrations were not constrained within 3$\sigma$.
	
	In Sec. \ref{sec:idialized}, we reported that the model always identifies H$_2$ as the dominant gas when the water concentration is greater than 10$^{-2}$. In the water world scenario (see Fig. \ref{fig:s3}), the framework suggests the water as the dominant gas. In that scenario, we included similar concentration for water and hydrogen ($\sim$0.5). The major difference between the water world scenario and the simple idealized scenario is the inclusion of more spectrally active gases. CH$_4$ and CO$_2$ add important details to the reflected spectrum that are useful to the Bayesian framework. The depth of the absorption features is indeed affected by the scale height of the atmosphere. If the atmosphere is H$_2$ or H$_2$O-dominated atmosphere, the mean molecular mass changes and therefore the atmospheric scale height. The inclusion of spectrally active gases helped to better converge towards the true values. 
	
	\edit1{Interestingly, the retrievals show broad constraints on the mixing ratio of N$_2$, a radiatively inactive gas (Fig. \ref{fig:s1}, \ref{fig:s2}, and \ref{fig:s3}). This is because the main role of N$_2$ is mainly to change the mean molecular mass. High value of N$_2$ would result in and heavy atmosphere and spectral features would be suppressed. This mechanism will define a possible upper limit. Low value of N$_2$ will instead create a lighter atmosphere where spectral absorption features are enhanced. This mechanism will help to define a lower limit. Finally, N$_2$, being a free parameter, is described by the modified priors introduced in this work. Extremely high value or very low value are unlikely, unless required by the data.}
	
	\section{Conclusion}
	\label{sec:conclusion}
	
	In this work, we presented the updated version of \exorelr\ for the direct imaging spectroscopy of sub-Neptune exoplanets. The update now allows the Bayesian framework to interpret reflected light spectra of temperate atmospheres with diverse bulk composition without any prior information about the composition. Small planets like those targeted in this study will not necessary have an H$_2$-dominated atmosphere, and therefore the dominant gas is unknown. We introduced the centered log ratio (CLR) as free parameter in our framework to allow compositional analysis. However, our attention is always on the more commonly used volume mixing ratio (VMR). To allow an unbiased exploration of VMR values we developed a new set of prior functions for the CLR. The shape of the distribution of the prior changes according to the number of gases to be fitted in the atmosphere. The more gases are introduced the flatter the prior distribution gets. 
	
	We tested the retrieval framework with idealized and realistic atmospheric scenarios. Initially, we simulated the simplest case: an atmosphere with only water and hydrogen. We simulated multiple instances of this scenario in which the water and the hydrogen concentration was changed. We reported that if the water VMR is comprised between 10${-5}$ and 10$^{-2}$, the water cloud present in the atmosphere is not opaque and photons are able to reach lower part of the atmosphere. \edit1{As a result, the cloud pressure and the water VMR can be retrieved simultaneously from the reflected light spectra.} On the contrary, if the water VMR is greater than 10$^{-2}$, the water cloud is optically thick and more photons are reflected. Increasing the VMR of water above 10$^{-2}$ does not change the modulation of the reflected spectrum resulting in spectral degeneracy.
	We showed that adding spectral active gases (e.g. CH$_4$ and CO$_2$) \edit1{helps break the degeneracy since now the dominant gas has spectral features and can thus be constrained from data.} In fact, the depth of the absorption features depend on the scale height of the atmosphere and the dominant gas plays a crucial role in determining the mean molecular mass value.
	
	With future space missions like HabEx \edit1{and LUVOIR}, temperate sub-Neptune planets will be at the forefront of the exoplanet exploration. Exploring models and implementing new statistic tools is therefore \edit1{an essential} step. In a future work, we will push forward the capabilities of our model to also implement the interpretation of super-Earths and Earth-like planets with a rocky surface that contributes to the overall light reflected by the planet (Reflected light of small planets II, Damiano \& Hu, in prep.).
	
	\section*{Acknowledgments}
	This work was supported in part by the NASA WFIRST Science Investigation Teams grant \#NNN16D016T. This research was carried out at the Jet Propulsion Laboratory, California Institute of Technology, under a contract with the National Aeronautics and Space Administration. 
	
	{	\small
		\bibliographystyle{apj}
		\bibliography{bib.bib}
	}
	
	\appendix
	\section{Idealized scenario}\label{sec:A_ideal}
	
\begin{figure*}[!h]
		\centering
		\includegraphics[angle=0, scale=0.40]{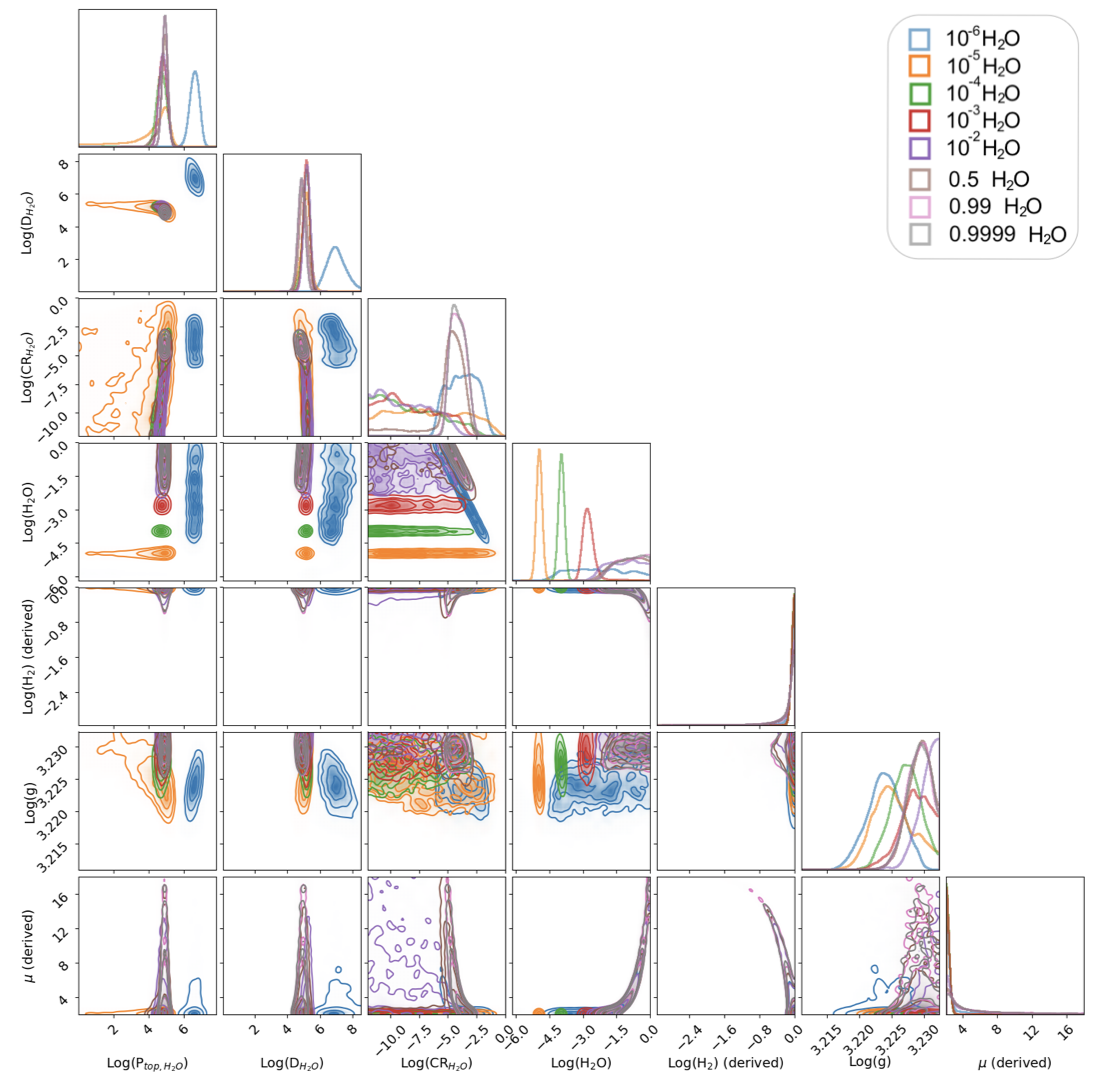}
		\caption{Full posterior distribution of the idealized cases considered in this work. The distribution shows that if the water abundance is $\geq10^{-2}$ the framework is not able to constraint the atmospheric composition. Moreover, if the mixing ratio of water is $\leq10^{-6}$, the forming water cloud is too thin to be statistically constrained. \label{fig:S_ideal}}
	\end{figure*}

	The idealized scenario is the simplest atmospheric scenario we used to test the capabilities of \exorelr. We test the model varying the abundance of water and hydrogen used for the scenarios. As described in Sec. \ref{sec:idialized}, the albedo spectra become degenerate when the water abundance is greater than $10^{-2}$ (Fig. \ref{fig:wh} and \ref{fig:S_ideal}). Moreover, the cloud position has been retrieved for all the cases but the case with water abundance of $10^{-6}$. In this case, the forming cloud is too thin to be statistically constrained.
	
	\newpage
	
	\section{Modified prior functions vs. constant prior in the CLR space}\label{sec:A_flat_vs_mod}
	
\begin{figure*}[!h]
		\centering
		\includegraphics[angle=0, scale=0.40]{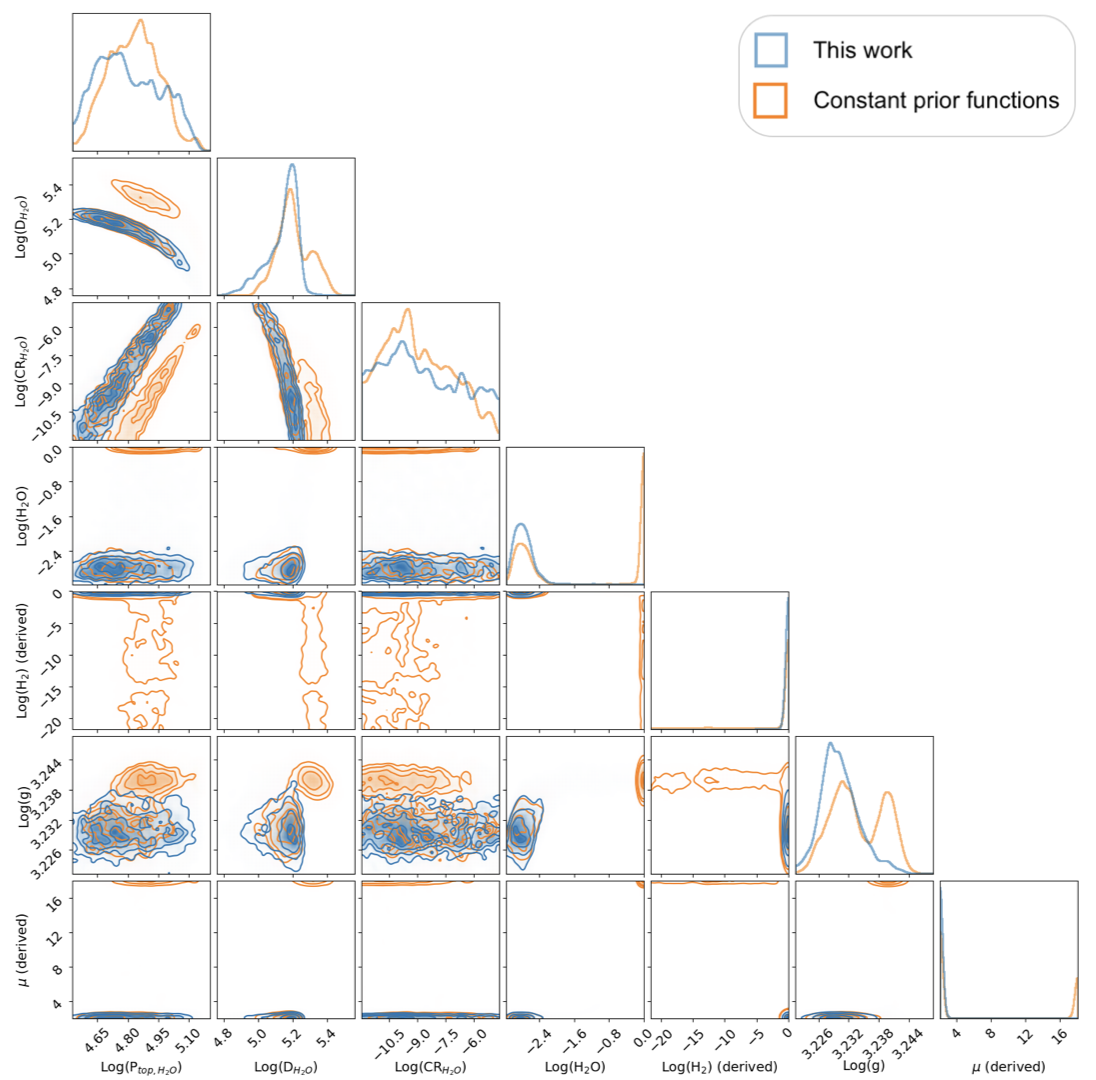}
		\caption{The full posterior distribution of the idealized case with water abundance of $10^{-3}$ in the case of constant or modified prior functions in the CLR space. The distribution shows that if the constant priors are used, a degeneracies are found in the water concentration and cloud properties. \label{fig:S_flat_vs_mod}}
	\end{figure*}

	In Sec. \ref{sec:priors} we introduced a new set of prior functions. The choice was justified by the non-linear transformation from the VMR space to the CLR space. In Fig.~\ref{fig:S_flat_vs_mod}, we compared the result of \exorelr when the constant or modified prior functions are used for one of the idealized scenario (i.e., the $10^{-3}$ H$_2$O case). As discussed in Sec. \ref{sec:priors}, when the constant priors are used the framework tends to statistically over-evaluate high abundance values of the gases (e.g., H$_2$O abundance in this case). Moreover, the retrieval using the constant prior results in a degenerate solution with a thicker cloud and a higher water mixing ratio below the cloud than the truth. The modified priors introduced in this work prevent the degenerate solution and	allow the framework to constraint all the free parameters without degeneracies.
	
	\section{Scenario 1: Pure water world}\label{sec:A1}
	
\begin{figure*}[!h]
		\centering
		\includegraphics[angle=0, scale=0.40]{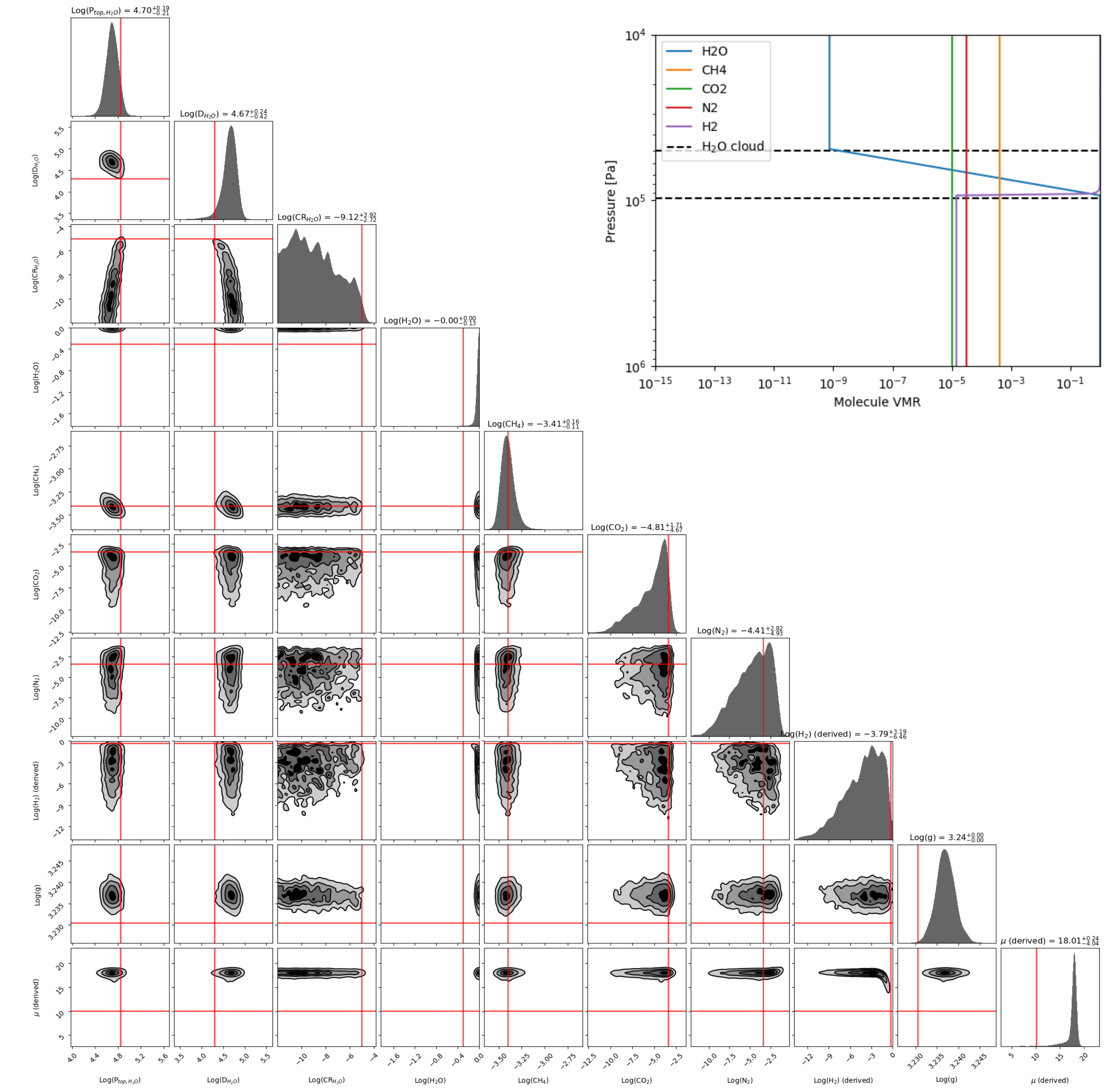}
		\caption{The full posterior distribution (corner plot) and the retrieved chemical profile of the scenario are shown. The red lines in the corner plot refer to the true value used to simulate the data. \label{fig:s1}}
	\end{figure*}

	When we refer to water world, we intend to describe a planet with high amount of water in the lower part of the atmosphere. We simulated a H$_2$O$-$dominated atmosphere with other absorbing gases. In Fig. \ref{fig:s1} we show the full posterior distribution and the retrieved chemical profile calculated by \exorelr. 
	
	\newpage
	
	\section{Scenario 2: Gas dwarf}\label{sec:A2}
	
\begin{figure*}[!h]
		\centering
		\includegraphics[angle=0, scale=0.40]{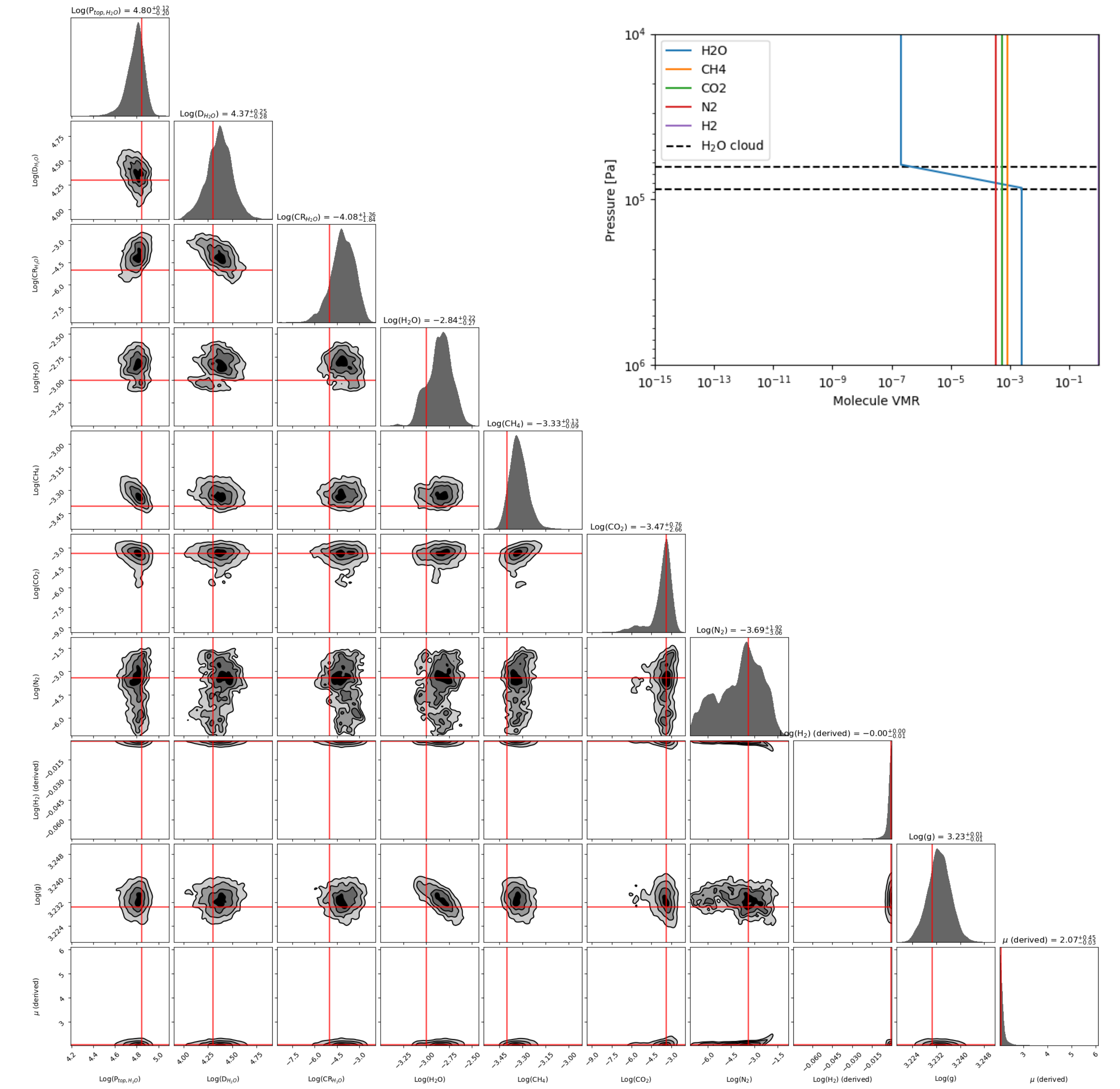}
		\caption{The full posterior distribution (corner plot) and the retrieved chemical profile of the scenario are shown. The red lines in the corner plot refer to the true value used to simulate the data.\label{fig:s2}}
	\end{figure*}

	The sub-Neptunes are a class of planets which have the dimension of a super-Earth but with a gaseous envelope. We simulated a H$_2-$dominated atmosphere with other absorbing gases. In Fig. \ref{fig:s2} we show the full posterior distribution and the retrieved chemical profile calculated by \exorelr.
	
	\newpage
	
	\section{Scenario 3: Water world with CO$_2$}\label{sec:A3}
	
\begin{figure*}[!h]
		\centering
		\includegraphics[angle=0, scale=0.40]{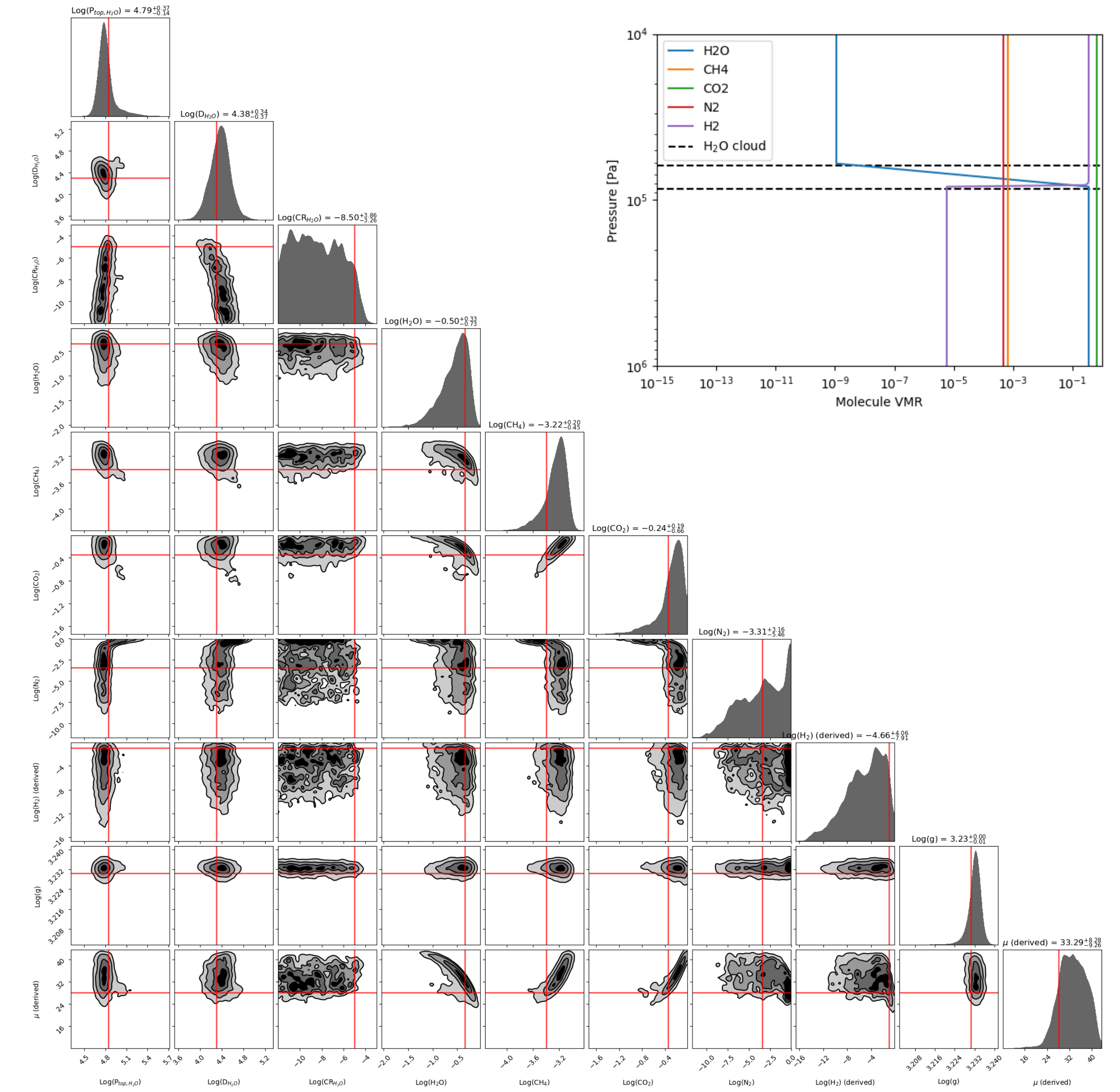}
		\caption{The full posterior distribution (corner plot) and the retrieved chemical profile of the scenario are shown. The red lines in the corner plot refer to the true value used to simulate the data.\label{fig:s3}}
	\end{figure*}

	We simulated a fictitious atmosphere without a clear single dominant gas. We rather chose to include H$_2$O and CO$_2$ with same volume mixing ratio ($\sim$0.5), and with other absorbing gases. In Fig. \ref{fig:s3} we show the full posterior distribution and the retrieved chemical profile calculated by \exorelr.
	
	\newpage
	
	\section{\textsc{MultiNest} implementation}
	
\begin{figure*}[!h]
	    \centering
	    \includegraphics[width=\textwidth]{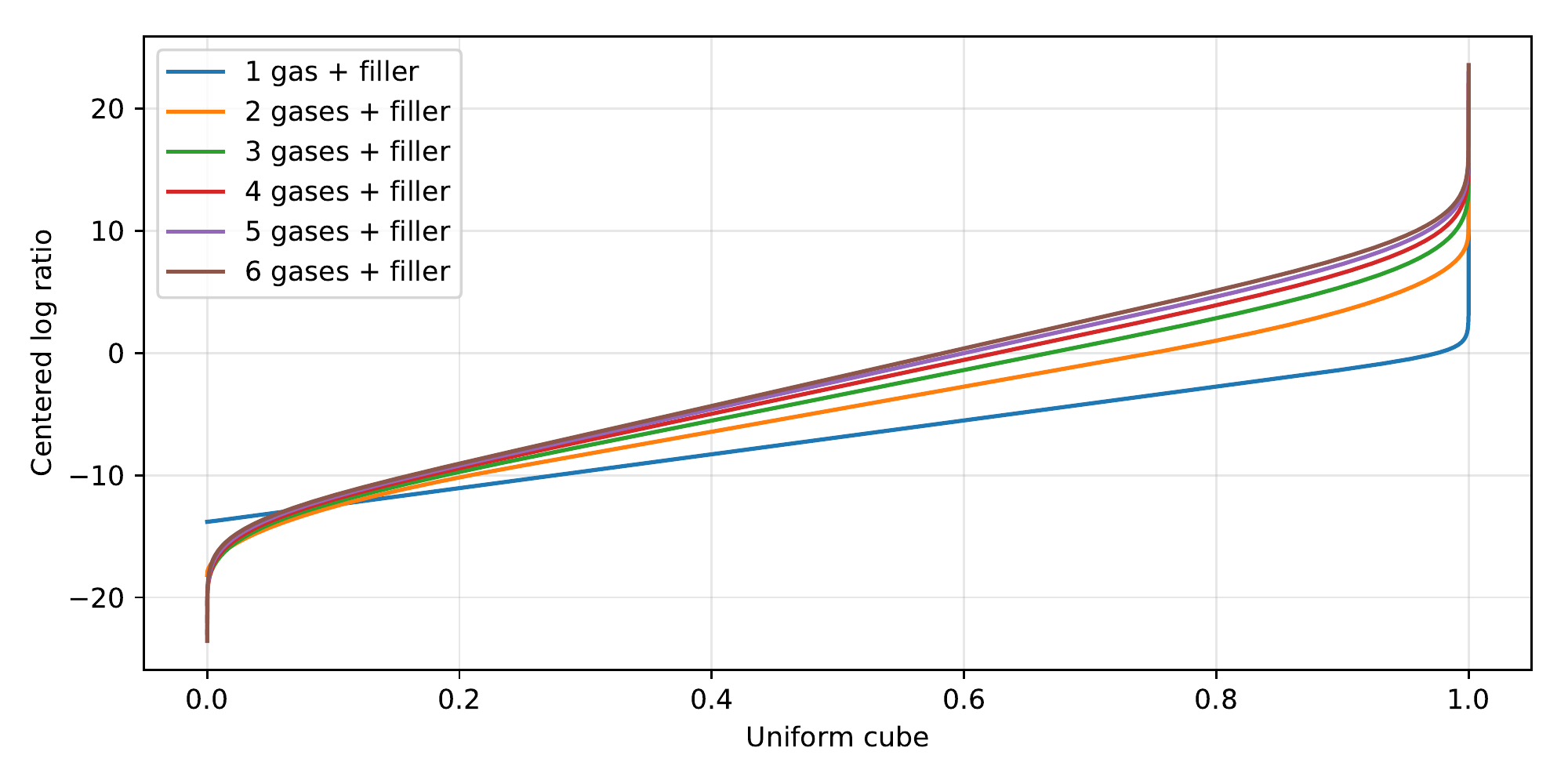}
	    \caption{Relation between the uniform cube defined inside \textsc{MultiNest} and the centered log ratio (CLR) defined in Sec. \ref{sec:priors} for different number of gases retrieved at the same time.}
	    \label{fig:priors1}
	\end{figure*}

	As we descibe in \cite{Damiano2020a}, we use \textsc{MultiNest} \citep{Buchner2014} as Bayesian engine. The prior functions in \textsc{MultiNest} are defined by transforming the uniform cube [0, 1] into the range of the physical quantities of interest. In the case of the work presented here, we need to calculate the inverse of the \textit{cumulative density function} (CDF) of the probability density function of the CLR obtained as described in Sec. \ref{sec:priors}. Fig. \ref{fig:priors1} shows the relation between the uniform cube and the CLR of the gases for different number of gases to be fit in the model. Note that once the number $n$ of gases has been chosen, the same prior function is used for all the $n$ gases.
	
\end{document}